\documentclass{article}
\usepackage{spconf}
\usepackage{dsfont}
\usepackage{mathdots}
\usepackage{color}
\usepackage{subfloat}
\usepackage{subfig}
\usepackage{graphicx}
\usepackage{amsmath}
\usepackage{epstopdf}
\usepackage{amsthm}
\usepackage{cite}
\usepackage{amssymb}
\usepackage{epsfig}
\usepackage{amsfonts}
\usepackage[]{algorithmic}
\usepackage[]{algorithm2e}
\usepackage{listings}
\usepackage[keeplastbox]{flushend}
\usepackage{fancybox}
\usepackage{epsfig}
\usepackage{mathtools}

\newcommand{\imj}{\mathsf{j}}
\usepackage{amsmath}

\title{Deep learning-based beam alignment in mmWave vehicular networks}
\name{Nitin Jonathan Myers, Yuyang Wang, Nuria Gonz{\'a}lez-Prelcic, and Robert W. Heath Jr. \thanks{This research was supported in part by the National Science Foundation under Grant numbers NSF-CNS-1731658 and NSF-CNS-1702800. }}
\address{Department of Electrical and Computer Engineering, The University of Texas at Austin.
  \\{Email :\small\tt \{nitinjmyers,yuywang,ngprelcic,rheath\}@utexas.edu }}

\begin{document}
\ninept
\maketitle
\begin{abstract}
Millimeter wave channels exhibit structure that allows beam alignment with fewer channel measurements than exhaustive beam search. From a compressed sensing (CS) perspective, the received channel measurements are usually obtained by multiplying a CS matrix with a sparse representation of the channel matrix. Due to the constraints imposed by analog processing, designing CS matrices that efficiently exploit the channel structure is, however, challenging. In this paper, we propose an end-to-end deep learning technique to design a structured CS matrix that is well suited to the underlying channel distribution, leveraging both sparsity and the particular spatial structure that appears in vehicular channels. The channel measurements acquired with the designed CS matrix are then used to predict the best beam for link configuration. Simulation results for vehicular communication channels indicate that our deep learning-based approach achieves better beam alignment than standard CS techniques that use the random phase shift-based design.
\end{abstract}
\section{Introduction}
\par Beam alignment in millimeter wave (mmWave) radios is a challenging problem due to the use of large antenna arrays and fewer radio frequency chains than antennas \cite{heathoverview}. Exhaustive search-based beam alignment results in a substantial overhead at mmWave \cite{11ad}. A possible approach to reduce this overhead is to use compressed sensing (CS)-based methods that acquire a lower dimensional channel representation \cite{csintro}. The use of random compressive channel projections together with algorithms that exploit sparsity of mmWave channels can achieve fast and accurate beam alignment \cite{cschest,javiCS}. Prior work has shown that structured compressive channel projections may result in better beam alignment than the use of random projections \cite{falp}.
\par Convolutional compressed sensing (CCS) is a structured CS technique in which the signal of interest is projected onto fewer circulant shifts of a known signal \cite{convCS}. We investigate 2D-CCS, i.e., CCS of 2D signals, as our focus is on planar antenna arrays. In 2D-CCS-based beam alignment, the transmitter (TX) applies fewer 2D-circulant shifts of a matrix to its antenna array for the receiver (RX) to acquire compressed channel measurements. Then, the best beam at the TX is estimated from the compressed measurements using optimization algorithms \cite{falp}. The matrix used in 2D-CCS, called the base matrix, determines the success of channel sparsity-aware beam alignment. In typical vehicular communication scenarios, channels exhibit structure beyond sparsity in the space of beam directions. For example, some beam directions may be more likely to be optimal than the others \cite{yuyang_access}. In such a case, the use of structured random CS matrices can result in better beam alignment than standard designs \cite{Anum_oob}. In this paper, we use deep learning as a tool to find a base matrix in 2D-CCS that is well suited to the channel prior. 
\par Prior work in imaging \cite{convcsnet} and beamforming \cite{alkhateebDL} has considered deep learning-based CS matrix optimization. In this paper, we propose a structured CS matrix optimization framework for the beam alignment problem. Our contributions are as follows. First, we show how 2D-CCS matrices with complex entries can be realized with real valued convolutional layers. Second, we propose a deep learning-based 2D-CCS matrix optimization procedure that accounts for the hardware constraints associated with radio frequency (RF) phase shifters. Last, we interpret the CS matrix optimized with deep learning using properties of the Fourier transform. The structured training in our 2D-CCS-based method results in fewer optimization parameters than the CS matrix optimization approach in \cite{alkhateebDL}.
\par  \textbf{Notation}$:$ $\mathbf{A}$ is a matrix, $\mathbf{a}$ is a column vector and $a, A$ denote scalars. The real and imaginary parts of $\mathbf{A}$ are denoted by $\mathbf{A}_{\mathrm{R}}$ and $\mathbf{A}_{\mathrm{I}}$. $A(k,\ell)$ denotes the entry of $\mathbf{A}$ in the $k^{\mathrm{th}}$ row and the ${\ell}^{\mathrm{th}}$ column. The $\ell^{\mathrm{th}}$ column of $\mathbf{A}$ is denoted by $\mathbf{A}(:,\ell)$. $|\mathbf{A}|$ is a matrix that contains the magnitude of the entries in $\mathbf{A}$. The Frobenius norm of $\mathbf{A}$ is $\Vert \mathbf{A} \Vert_{\mathrm{F}}$. The inner product of two matrices $\mathbf{A}$ and $\mathbf{B}$ is defined as $\langle \mathbf{A},\mathbf{B}\rangle =\sum_{k,\ell}A(k,\ell)B(k,\ell)$. The matrix $[\mathbf{A};\mathbf{B}]$ is obtained by vertically stacking $\mathbf{A}$ and $\mathbf{B}$. $\mathbf{U}_N$ is an $N\times N$ unitary discrete Fourier transform (DFT) matrix. $\mathsf{j}=\sqrt{-1}$.  
\section{Beam alignment with convolutional CS}
\begin{figure}[h!]
\centering
\includegraphics[trim=0.5cm 0.2cm 0.5cm 0cm,clip=true,width=0.4\textwidth]{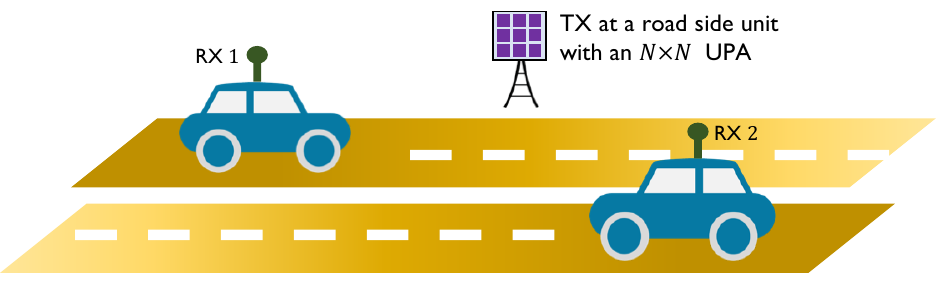}
\vspace{0.5mm}
  \caption{An illustration of a vehicular communication scenario considered in this paper. The RSU serves vehicles on the two lanes.}
  \label{fig:System_setting}
\vspace{-4mm}  
\end{figure}

We consider a narrowband system with an $N \times N$ uniform planar array (UPA) at the TX and a single antenna at the RX. The single antenna assumption at the RX is made for simplicity; an extension of our approach to receivers with multiple antennas will be considered in our future work. The TX is equipped with a $q$-bit phased antenna array that uses a single RF chain and is mounted on a road side unit (RSU). The RSU serves an RX which is mounted on top of a vehicle, as illustrated in Fig.~\ref{fig:System_setting}. We use $\mathbf{H} \in \mathbb{C}^{N \times N}$ to denote the channel matrix between the UPA at the TX and the single antenna RX. Note that the $(r,c)^{\mathrm{th}}$ entry of $\mathbf{H}$ represents the channel coefficient between the $(r,c)^{\mathrm{th}}$ antenna at the TX and the RX antenna. We consider both line-of-sight (LoS) and non-LoS channels in our simulations. We use $\mathbb{Q}_q$ to denote the set of possible phase shifts in the analog beamforming network, i.e., $\mathbb{Q}_q=\{e^{\imj 2 \pi b/ 2^q}/ N : b \in \{1,2,\cdots 2^q\}\}$. In the $m^{\mathrm{th}}$ beam training slot, the TX applies the phase shift matrix $\mathbf{P}[m] \in \mathbb{Q}^{N \times N}_q$ to its phased array. The channel measurement received by the RX for a unit pilot symbol is then
\begin{equation}
\label{eq:sysmodel}
y[m]=\langle \mathbf{H}, \mathbf{P}[m] \rangle +v[m],
\end{equation}
where $v[m] \in \mathbb{C}$ denotes additive white Gaussian noise (AWGN). We define $\mathcal{F}$ as the standard 2D-DFT codebook for the UPA-based TX. The transmit beam alignment problem is to estimate a beamformer $\mathbf{P}_{\mathrm{BF}} \in \mathcal{F}$ that maximizes the inner product $|\langle \mathbf{H},\mathbf{P}\rangle|$. 
\par A straightforward approach for beam alignment is to first estimate all the $N^2$ coefficients of $\mathbf{H}$ using scalar projections of the form in \eqref{eq:sysmodel}. Then, beam alignment can be performed using the estimated channel. Estimating the $N \times N$ channel matrix, however, can result in a significant overhead as the channel dimension is large in typical mmWave settings. Due to the structure in mmWave vehicular channels, it may be possible to estimate the best beam without explicit channel estimation. For example, prior work has shown that beam alignment can be performed with just $\mathcal{O}(\mathrm{log}\, N)$ random phase shift-based compressed measurements of a sparse channel \cite{javiCS}. Although random phase shift-based CS performs well, the question is if it is possible to construct a CS matrix that is better matched to the channel prior in vehicular settings.
\begin{figure}[htbp]
\vspace{-6mm}
\centering
\subfloat[Beamspace prior (log scale)]{\includegraphics[trim=0cm 0cm 1.5cm 0.5cm,clip=true, width=4.25cm, height=3.8cm]{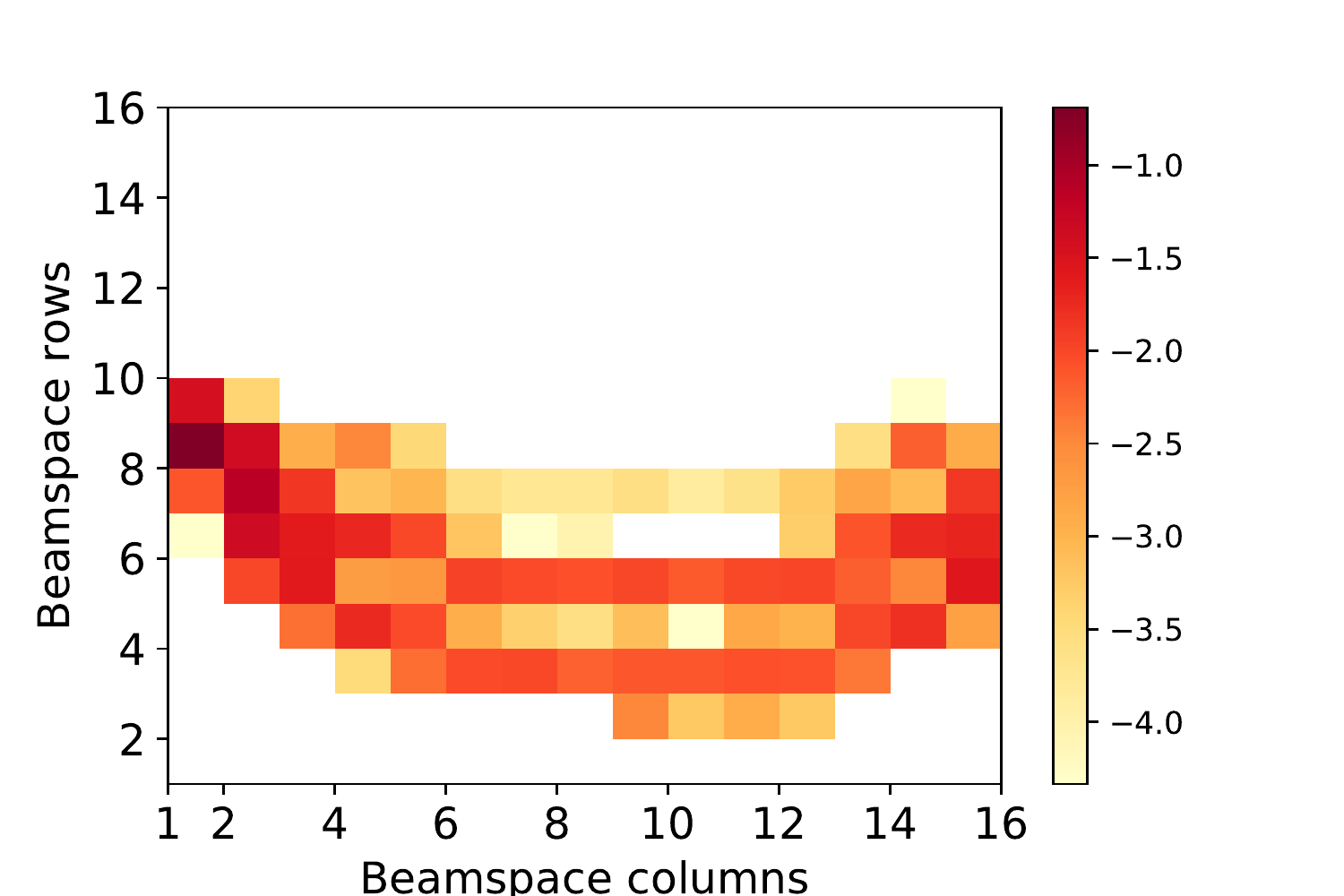}\label{fig:beamspace_prior}}
\subfloat[Beampattern of a random matrix]{\includegraphics[trim=0cm 0cm 1.5cm 0.5cm,clip=true, width=4.25cm, height=3.8cm]{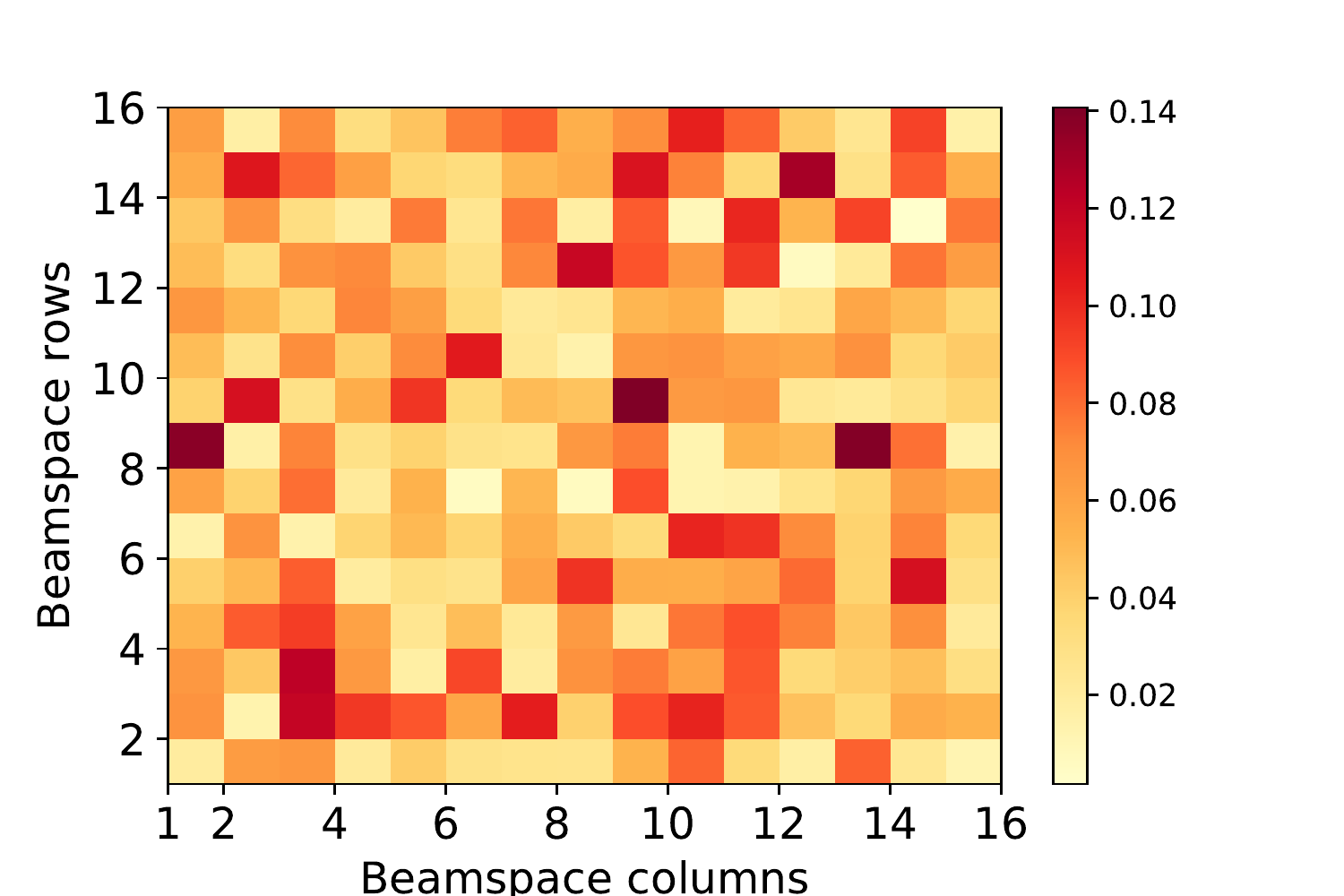}\label{fig:random_beam}}
\vspace{1mm}
\caption{ \small For the vehicular communication scenario in our simulations, the beamspace prior is non-zero over a small set in the space of 2D-DFT directions. Random phase shift-based CS uses quasi-omnidirectional beams and does not exploit such an information.  
\normalsize}
\vspace{-2mm}
\end{figure}
\par We now discuss the additional channel structure present in a typical mmWave vehicular communication scenario, such as the one shown in Fig.~\ref{fig:System_setting}. For an ensemble of receivers within the coverage of an RSU, we show the probability distribution of the best transmit beams in Fig. \ref{fig:beamspace_prior}. The $(i,j)^{\mathrm{th}}$ entry of the beamspace prior in Fig. \ref{fig:beamspace_prior} is the probability that the $(i,j)^{\mathrm{th}}$ 2D-DFT beam, i.e., $\mathbf{U}_N(:,i)\mathbf{U}^T_N(:,j)$, is optimal. In our simulation scenario, the beamspace prior is concentrated over two strips on an $N \times N$ grid of 2D-DFT angles. These two strips correspond to directional beams that point along the vehicle moving trajectories associated with two different lanes shown in Fig. \ref{fig:System_setting}. Compressed sensing using random phase shift matrices, however, does not exploit the fact that the beamspace probability is concentrated in a small support set. This is because the beam associated with a random phase shift matrix is quasi-omnidirectional with a high probability; an example of such a beam is shown in Fig. \ref{fig:random_beam}. Sensing with a quasi-omnidirectional pattern is inefficient in vehicular applications, since the vehicle locations are constrained to the road lanes, and many angular directions can be discarded apriori.
\subsection{Motivation to use 2D-CCS over standard CS}
\par We first describe the measurement model in 2D-CCS when the TX uses $\mathbf{P} \in \mathbb{Q}^{N \times N}_q$ as the base matrix \cite{falp}. In 2D-CCS, the TX applies $M$ distinct 2D-circulant shifts of $\mathbf{P}$ to its phased array for the RX to acquire channel measurements. To explain 2D-CCS, we consider a case where the TX applies all the $N^2$ possible 2D-circulant shifts of $\mathbf{P}$ to its antenna array. In this case, $M=N^2$ and the collection of measurements acquired by the RX is the 2D-circular cross-correlation between $\mathbf{H}$ and $\mathbf{P}$. We denote this circular cross-correlation as
\begin{equation}
\label{eq:defn_G}
\mathbf{G}=\mathbf{H}\star \mathbf{P}.
\end{equation}
The $(r,c)^{\mathrm{th}}$ entry of $\mathbf{G}$ can be expressed as 
\begin{equation}
G(r,c)=\sum_{k,\ell}H(k,\ell)  P\left((k-r)\, \mathrm{mod} \,N,(\ell-c)\, \mathrm{mod}\, N \right).
\end{equation}
Acquiring all the $N^2$ entries of $\mathbf{G}$, however, results in a training overhead that is comparable to exhaustive search. In 2D-CCS, the TX applies $M \ll N^2$ random 2D-circulant shifts of $\mathbf{P}$ to reduce the overhead. We use $\Omega$ to denote an ordered set of $M$ circulant shifts used by the TX. With the circulant shift-based training, the RX acquires a subsampled version of $\mathbf{G}$ at the ordered locations in $\Omega$.  For example, $\Omega=\{(0,1), (1,2),(2,1)\}$ is one possible set for $M=3$. In this case, the RX acquires $G(0,1),G(1,2)$ and $G(2,1)$. We define $\mathcal{P}_{\Omega}(\mathbf{G})$ as the vector containing the entries of $\mathbf{G}$ at the locations in $\Omega$. The compressed channel measurement vector in 2D-CCS is then
\begin{equation}
\label{eq:Subsamp_G}
\mathbf{y}=\mathcal{P}_{\Omega}(\mathbf{G})+\mathbf{v}.
\end{equation}
The measurements are called convolutional channel measurements as the subsampled cross-correlation operation in \eqref{eq:Subsamp_G} can be realized using subsampled convolution. For a well designed base matrix $\mathbf{P}$ in 2D-CCS, optimization algorithms can estimate the best beamspace direction from the compressed channel measurements in \eqref{eq:Subsamp_G} even when $M \ll N^2$ \cite{falp}.  
\par We explain the efficiency of 2D-CCS in exploiting the beamspace prior with the Fourier transform. We define $\mathbf{X}$ as the 2D-DFT of the channel $\mathbf{H}$ and $\mathbf{Z}[m]$ as the 2D-DFT of the phase shift matrix $\mathbf{P}[m]$. The matrix $\mathbf{X}$ is called the beamspace and the prior associated with the maximizer of $|\mathbf{X}|$ is shown in Fig. \ref{fig:beamspace_prior}. This prior is also the probability distribution of the best beam within the 2D-DFT codebook. To explain the efficiency of 2D-CCS, we consider an example where $\mathbf{P}[1]$ is set to a given base matrix, i.e., $\mathbf{P}[1]=\mathbf{P}$. The first channel measurement is then $\langle \mathbf{H}, \mathbf{P} \rangle$. Due to the unitary nature of the 2D-DFT, the inner product between $\mathbf{H}$ and $ \mathbf{P}$ can also be expressed as the inner product of their 2D-DFTs, i.e., $\langle \mathbf{X}, \mathbf{Z}[1] \rangle$. It can be observed from Fig. \ref{fig:beamspace_prior} that the information required for beam alignment is encoded in fewer coefficients of $\mathbf{X}$, i.e., the entries of $\mathbf{X}$ along the two strips. To extract this information from $\langle \mathbf{X}, \mathbf{Z}[1] \rangle$ in a way that is robust to noise, $\mathbf{Z}[1]$ must have a large amplitude at the locations along the two strips. A base matrix $\mathbf{P}$ whose 2D-DFT satisfies such a property can be used to efficiently acquire features or measurements of the sparse channel.  
\par An interesting property of 2D-CCS is that the matrices used to obtain channel projections have the same 2D-DFT magnitude as the base matrix, i.e., $|\mathbf{Z}[m]|=|\mathbf{Z}[1]| \, \forall m$. The observation follows from the fact that 2D-circulant shifts over a matrix do not change the magnitude of its 2D-DFT \cite{imageprocess}. Using this property, it can be observed that a careful design of the base matrix ensures that all the matrices $\{\mathbf{Z}[m]\}^M_{m=1}$ have a large amplitude at the desired locations. Designing a base matrix that is best suited to the channel prior, however, can be challenging when $M\ll N^2$. In Section \ref{sec:DL_main}, we discuss how deep learning can be used to solve the base matrix design problem for efficient 2D-CCS. We assume that the subsampling set $\Omega$ is chosen at random and is fixed throughout optimization. Optimizing the subsampling set is an interesting direction for future work.
\section{Base matrix design with deep learning} \label{sec:DL_main}
\begin{figure*}[h!]
\centering
\includegraphics[trim=3.2cm -0.2cm 3cm 1cm, width=0.7\textwidth]{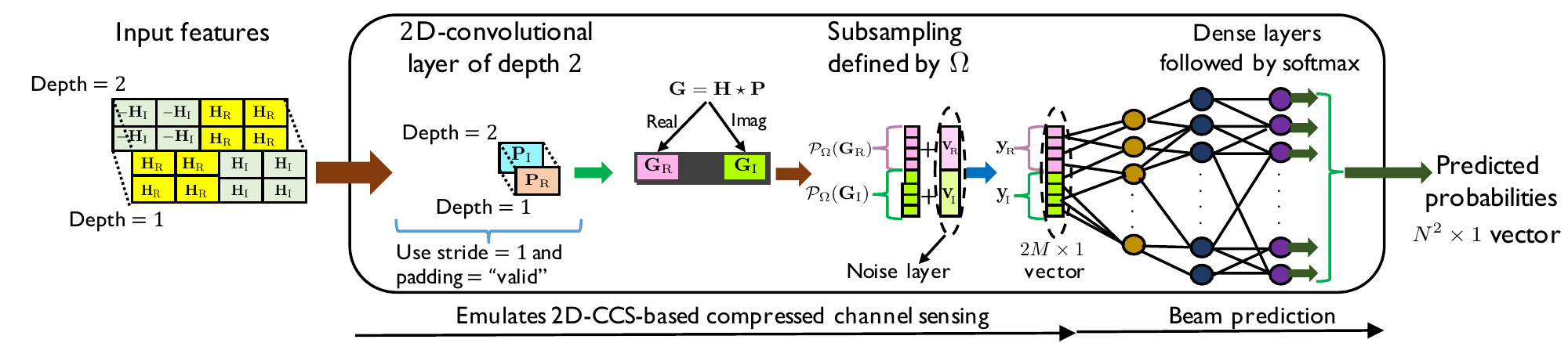}
\vspace{1mm}
\caption{\small Channel measurements in 2D-CCS are realized using convolutional filters $\mathbf{P}_{\mathrm{R}}$ and $\mathbf{P}_{\mathrm{I}}$. Using end-to-end learning, the base matrix in 2D-CCS, i.e.,  $\mathbf{P}=\mathbf{P}_{\mathrm{R}}+\imj \mathbf{P}_{\mathrm{I}}$, is optimized to maximize the probability of beam alignment using the compressed channel measurements.}
\normalsize
\vspace{-1.75mm}
  \label{fig:2dccs_illus}
\end{figure*}
\par We explain the proposed deep neural network architecture using Fig. \ref{fig:2dccs_illus}. The first part of our deep neural network contains a convolutional layer to emulate 2D-CCS-based compressed channel sensing in \eqref{eq:Subsamp_G}. The weights of the filters in the convolutional layer model the real and imaginary components of the base matrix $\mathbf{P}$ in 2D-CCS. The second part of the network is a cascade of fully connected layers that predicts the best beam from the compressed channel measurements. Through end-to-end learning, our network learns a base matrix $\mathbf{P}$ and the weights of the fully connected layers that maximize the beam alignment probability. In this section, we describe the key components of our network and the training procedure; our implementation is available online on GitHub \cite{Code_deep_learning}. 
\subsection{2D-CCS using real valued convolutional layers}\label{sec:2dccs_implem}
\par We explain how to implement the circular cross-correlation in \eqref{eq:defn_G} using the example of a real matrix $\mathbf{H}_{\mathrm{R}} \star \mathbf{P}_{\mathrm{R}}$. To realize a circulant structure in the correlation, we define a matrix $\mathbf{A}_{\mathrm{R,pad}}=[\mathbf{A}_{\mathrm{R}},\mathbf{A}_{\mathrm{R}};\mathbf{A}_{\mathrm{R}},\mathbf{A}_{\mathrm{R}}]$. Note that $\mathbf{H}_{\mathrm{R,pad}}$ is a $2N \times 2N$ matrix. Now, applying the convolutional filter $\mathbf{P}_{\mathrm{R}}$ over $\mathbf{H}_{\mathrm{R,pad}}$, in a valid-padding mode with a stride of $1$, results in an $(N+1)\times (N+1)$ matrix. The circular cross-correlation $\mathbf{H}_{\mathrm{R}} \star \mathbf{P}_{\mathrm{R}}$ is obtained by simply deleting the last row and the last column of the convolved output. 
\par Our network obtains the complex valued circular cross-correlation by computing the real and imaginary components of $\mathbf{G}$ in \eqref{eq:defn_G}. As $\mathbf{P}=\mathbf{P}_{\mathrm{R}}+\imj \mathbf{P}_{\mathrm{I}}$, the real and imaginary components of $\mathbf{G}$ are
\begin{align}
\label{eq:G_Rdefn}
\mathbf{G}_{\mathrm{R}}&=\mathbf{H}_{\mathrm{R}} \star \mathbf{P}_{\mathrm{R}}-\mathbf{H}_{\mathrm{I}} \star \mathbf{P}_{\mathrm{I}},\,\,\, \mathrm{and}\\
\label{eq:G_Idefn}
\mathbf{G}_{\mathrm{I}}&=\mathbf{H}_{\mathrm{I}} \star \mathbf{P}_{\mathrm{R}}+\mathbf{H}_{\mathrm{R}} \star \mathbf{P}_{\mathrm{I}}.
\end{align}
To realize \eqref{eq:G_Rdefn} and \eqref{eq:G_Idefn}, a $2N\times 4N \times 2$ real valued tensor is first constructed from the complex channel $\mathbf{H}$. The first slice of this tensor is the $2N\times 4N $ matrix $[\mathbf{H}_{\mathrm{R,pad}},\mathbf{H}_{\mathrm{I,pad}}]$ and the second slice is $[-\mathbf{H}_{\mathrm{I,pad}},\mathbf{H}_{\mathrm{R,pad}}]$, as shown in Fig. \ref{fig:2dccs_illus}. Then, an $N\times N \times 2$ convolutional filter with $\mathbf{P}_{\mathrm{R}}$ and $\mathbf{P}_{\mathrm{I}}$ as the filter weights, is applied over the input channel tensor. The filtering is performed in a valid-padding mode with a stride of $1$ to result in an $(N+1)\times (3N+1)$ matrix. Finally, $\mathbf{G}_{\mathrm{R}}$ and $\mathbf{G}_{\mathrm{I}}$ are $N \times N$ matrices which begin at the $(0,0)$ and $(2N,0)$ coordinates of the $(N+1)\times (3N+1)$ matrix.
\par The channel measurements in \eqref{eq:Subsamp_G} can be split into real and imaginary components as $\mathbf{y}_{\mathrm{R}}=\mathcal{P}_{\Omega}(\mathbf{G}_{\mathrm{R}})+\mathbf{v}_{\mathrm{R}}$ and $\mathbf{y}_{\mathrm{I}}=\mathcal{P}_{\Omega}(\mathbf{G}_{\mathrm{I}})+\mathbf{v}_{\mathrm{I}}$. It is important to note that the subsampling set in \eqref{eq:Subsamp_G}, i.e., $\Omega$,  is same for both the real and the imaginary components of $\mathbf{y}$. We implement such a subsampling by using an identical dropout technique for both $\mathbf{G}_{\mathrm{R}}$ and $\mathbf{G}_{\mathrm{I}}$. Our dropout procedure discards $N^2-M$ entries of $\mathbf{G}_{\mathrm{R}}$ and $\mathbf{G}_{\mathrm{I}}$ at the same locations which are chosen at random. The subsampled features, i.e., $\mathcal{P}_{\Omega}(\mathbf{G}_{\mathrm{R}})$ and $\mathcal{P}_{\Omega}(\mathbf{G}_{\mathrm{I}})$, are perturbed by AWGN to emulate the compressed sensing model in \eqref{eq:Subsamp_G}. 
\subsection{End-to-end learning for beam prediction} 
The subsampled channel feature vector of dimension $2M \times 1$ is fed into a beam prediction network that is a cascade of fully connected layers. ReLU activation is used at all the fully connected layers and a softmax is used at the output layer. In this paper, beam alignment is performed using a 2D-DFT codebook that has $N^2$ elements. Each codebook element is modeled by a class and the output of the neural network is an $N^2 \times 1$ vector that contains the predicted class probabilities.  
\par The network is trained using the restructured channels, i.e., a collection of $2N \times 4N \times 2$ tensors defined in Section \ref{sec:2dccs_implem}, and the best beam class indices corresponding to the channels. The deep neural network is trained by minimizing the cross entropy between the predicted class probabilities and the one-hot encoded vectors corresponding to the best beam index. It is important to note that the convolutional filters $\mathbf{P}_{\mathrm{R}}$ and $\mathbf{P}_{\mathrm{I}}$ are part of the neural network and are optimized during training. After successful training, the beam prediction component of the neural network is expected to output the best beam direction from the $2M \times 1$ feature vector which represents the compressed channel measurements. 
\subsection{Weight quantization and implementation in hardware} 
The filter weights $\mathbf{P}_{\mathrm{R}}$ and $\mathbf{P}_{\mathrm{I}}$ trained with end-to-end learning may not be hardware compatible, i.e., the corresponding base matrix $\mathbf{P}=\mathbf{P}_{\mathrm{R}}+ \imj \mathbf{P}_{\mathrm{I}}$ may not belong to $\mathbb{Q}^{N \times N}_q$. To this end, we use an approach similar to weight quantization and retraining in \cite{han2015deep}. We define $\mathrm{phase}_q(w)$ as a function that returns the phase of $w$ rounded to a nearest integer multiple of $2 \pi /2^q$. The $q$-bit phase quantized version of  $\mathbf{P}$ is defined as $\mathbf{P}_{\mathrm{quant}}=\mathrm{exp}(\imj\, \mathrm{phase}_q(\mathbf{P}))/N$. The optimized filter weights $\mathbf{P}_{\mathrm{R}}$ and $\mathbf{P}_{\mathrm{I}}$ are replaced by the real and imaginary parts of $\mathbf{P}_{\mathrm{quant}}$. Replacing the filter weights in the convolutional layer, however, can reduce the accuracy of the trained neural network. To compensate for the performance loss due to quantization, the fully connected layers following the convolutional layers are retrained. It is important to note that the  convolutional layer containing $\mathbf{P}_{\mathrm{quant}}$ is not updated during retraining.
\par Now, we explain how the trained network can be used for beam alignment in mmWave phased arrays. First, the TX applies $M$ circulant shifts of $\mathbf{P}_{\mathrm{quant}}$ to its phased array, according to the subsampling set $\Omega$. For each phase shift matrix applied at the TX, the RX acquires a channel measurement defined by \eqref{eq:sysmodel}. Therefore, the RX obtains a subsampled version of $\mathbf{G}=\mathbf{H} \star \mathbf{P}_{\mathrm{quant}}$ according to the model in \eqref{eq:Subsamp_G}. The received measurements are then reshaped into a vector $[\mathbf{y}_{\mathrm{R}};\mathbf{y}_{\mathrm{I}}]$ which is fed into the trained fully connected layers for beam prediction. The RX feedbacks the index of this beam and the TX applies the corresponding 2D-DFT codebook element. 
\section{Simulations}
\par We consider a vehicular communication scenario in which the RSU is placed at a height of $5 \, \mathrm{m}$. The TX at the RSU is equipped with a $16 \times 16 $ UPA with $3$-bit phase shifters, i.e., $N=16$ and $q=3$. The carrier frequency is set as $28\, \mathrm{GHz}$ and the operating bandwidth is $100\, \mathrm{MHz}$. In our simulation scenario, vehicles move along two lanes that are at a distance of $4\, \mathrm{m}$ and $7 \, \mathrm{m}$ from the foot of the RSU. We use Wireless Insite \cite{wireless_insite}, a ray tracing simulator, to obtain channel matrices between the TX and the receivers which are mounted on top of the vehicles. The beamspace prior corresponding to this scenario is shown in Fig. \ref{fig:beamspace_prior}. Note that our dataset contains both LoS and NLoS channels. The NLoS scenario  occurs when the link between the RSU and a vehicle is blocked by tall trucks.
\par The convolutional layer in our network consists of a single filter of dimensions $16 \times 16 \times 2$. This filter is used to optimize the base matrix in 2D-CCS. To emulate $M$ number of 2D-CCS-based channel measurements, the convolved output is subsampled to a $2M \times 1$ feature vector. This vector is then fed into a cascade of four fully connected layers with output dimensions of $80$, $256$, $512$ and $256$. For the beam alignment problem, a reasonable network is one that is invariant to the scaling of the channel. 
The bias at all the layers is forced to zero to ensure that the network satisfies the scale invariance property. We use about $20,000$ channels to train the network and $5,000$ other channels for testing. Every channel in the training set is scaled so that $\Vert \mathbf{H} \Vert_{\mathrm{F}}=N$. For the test set, however, we use a common scaling for all the channels such that $\mathbb{E}[\Vert \mathbf{H} \Vert^2_{\mathrm{F}}]=N^2$. Here, $\mathbb{E}[\cdot]$ denotes the average across the test channels. The channels are then restructured according to the procedure in Section \ref{sec:2dccs_implem}.
\begin{figure}[htbp]
\vspace{-6mm}
\centering
\subfloat[Beam pattern before quantization]{\includegraphics[trim=0cm 0cm 1.5cm 0.5cm,clip=true, width=4.25cm, height=3.7cm]{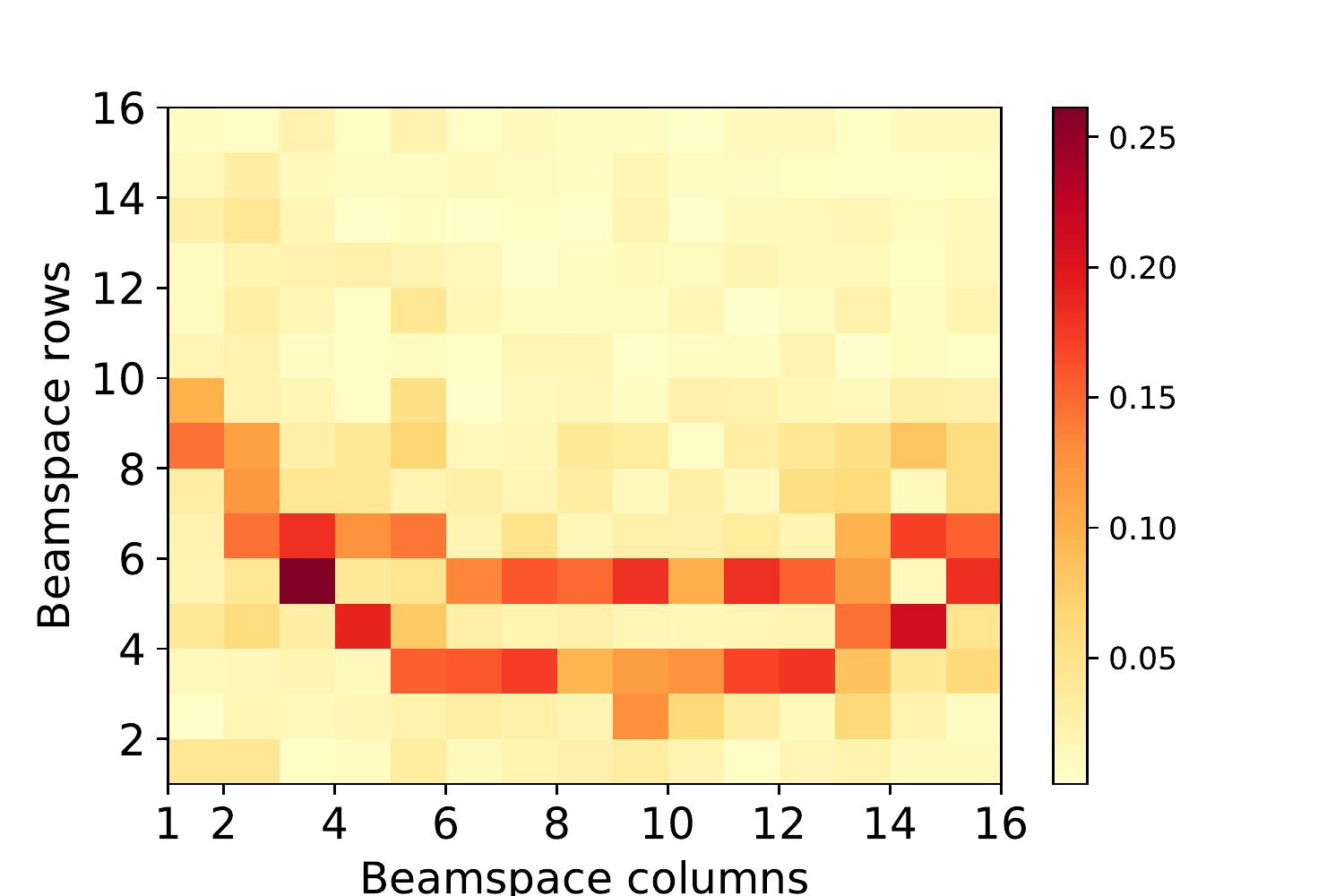}\label{fig:beam_prequant}}
\subfloat[Beam pattern after quantization]{\includegraphics[trim=0cm 0cm 1.5cm 0.5cm,clip=true, width=4.25cm, height=3.7cm]{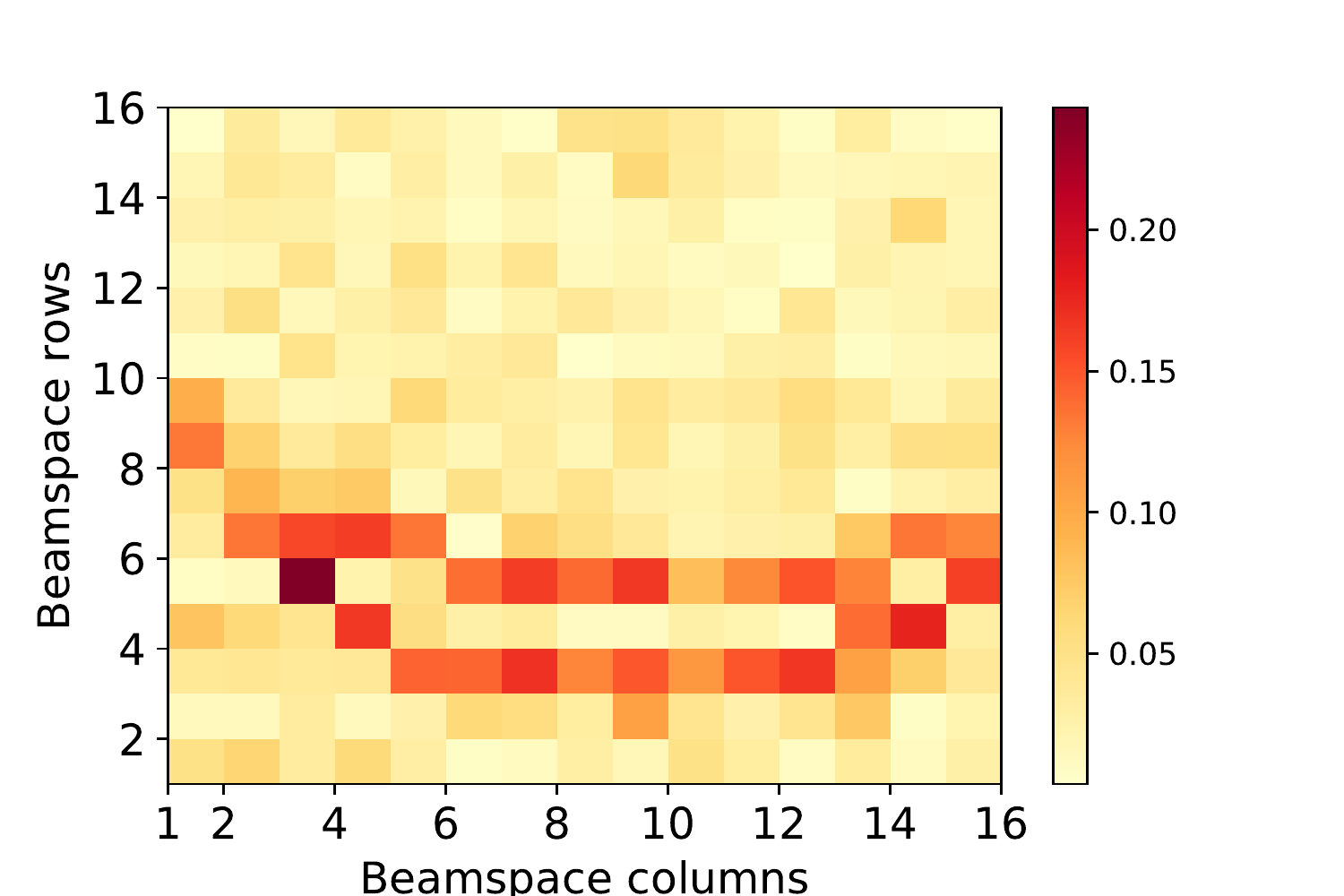}\label{fig:beam_postquant}}
\vspace{1mm}
\caption{ \small The plot shows the 2D-DFT magnitude of the base matrices optimized with our procedure for $M=10$. Weight quantization slightly perturbs the beam pattern. The beams in optimized 2D-CCS radiate power along the directions that are more likely to be optimal.
\normalsize}
\end{figure}
\par We explain how our network is trained and provide insights into the optimized base matrix. For $M$ channel measurements, our method first samples $M$ distinct 2D-integer coordinates from an $N \times N$ grid at random and constructs the subsampling set $\Omega$. The $N \times N$ integer grid corresponds to the support of $\mathbf{G}$ in \eqref{eq:defn_G}. Then, the proposed deep neural network is trained for $30$ epochs with the restructured channels and the associated class labels from the training set. A good base matrix is one whose beam pattern has a large magnitude along the directions that are more likely to be optimal. A discrete version of this beam pattern is the 2D-DFT of the base matrix. For the scenario corresponding to Fig. \ref{fig:beamspace_prior}, the 2D-DFT of the optimized base matrix derived after training is shown in Fig. \ref{fig:beam_prequant}. A $3$-bit phase quantization is performed over the optimized base matrix to obtain $\mathbf{P}_{\mathrm{quant}}$; the 2D-DFT of $\mathbf{P}_{\mathrm{quant}}$ is shown in Fig. \ref{fig:beam_postquant}. It can be observed that the beam pattern corresponding to $\mathbf{P}_{\mathrm{quant}}$ is well matched to the beamspace prior in Fig. \ref{fig:beamspace_prior}. As quantization can lower the beam prediction accuracy of the trained network, the four fully connected layers in our network are retrained with $30$ epochs.
 \begin{figure}[h!]
\centering
\includegraphics[trim=0.3cm -0.4cm 1cm 1cm,clip=true,width=0.4\textwidth]{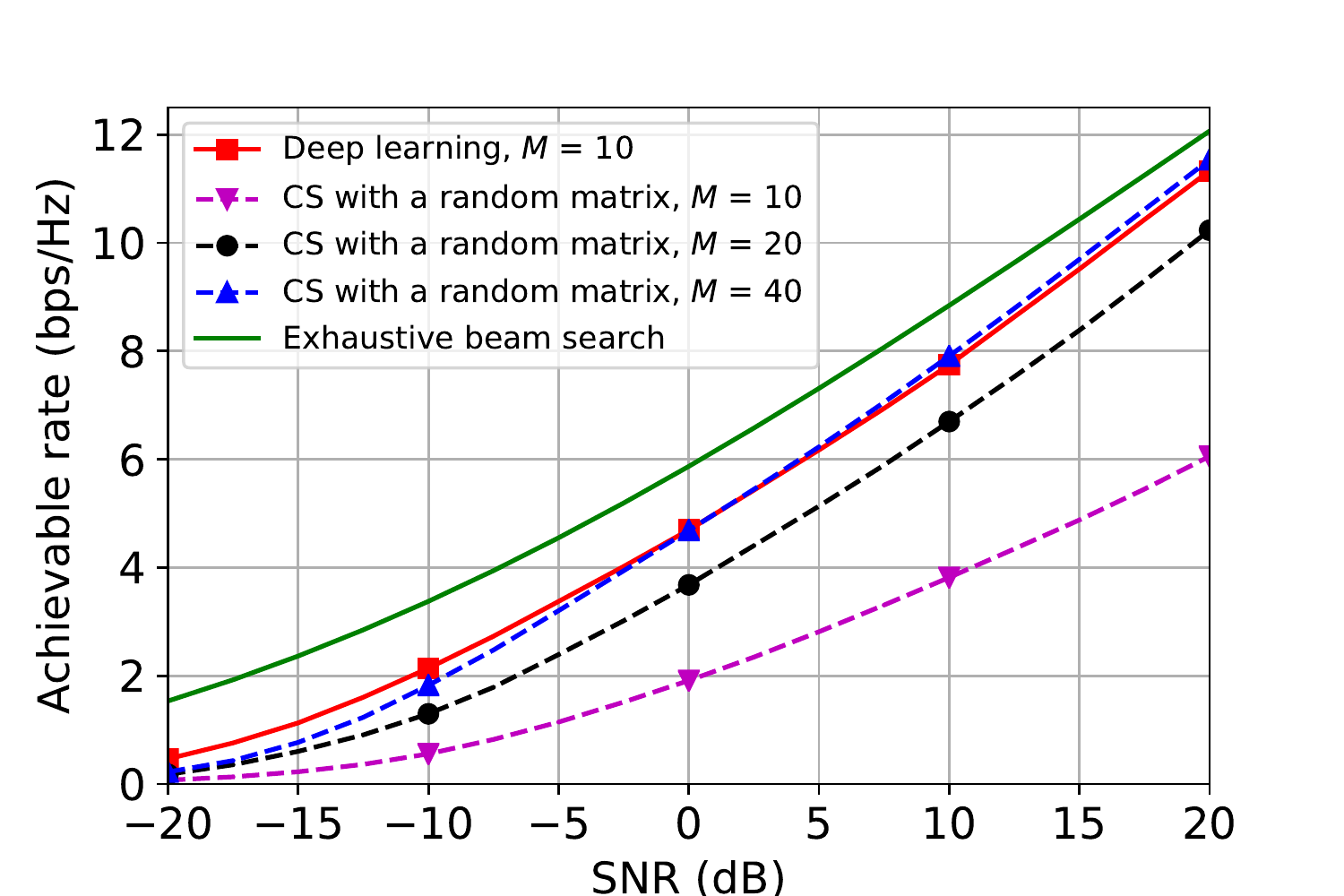}
  \caption{Deep learning-based beam alignment with an optimized base matrix performs better than standard CS with a random phase shift-based design. The proposed approach can reduce the training overhead by $4 \times$ when compared 
 to standard CS.}
  \label{fig:Rate_vs_SNR}
\end{figure}
\par We evaluate the proposed deep learning-based 2D-CCS technique in terms of the achievable rate over the test dataset. The $2M \times 1$ compressed channel measurement vector after subsampled convolution with $\mathbf{P}_{\mathrm{quant}}$ is perturbed by AWGN of variance $\sigma^2/2$. The SNR observed at the RX when the TX uses a quasi-omnidirectional pattern is defined as $1/\sigma^2$. As we focus on beam alignment using the 2D-DFT codebook, the SNR after beamforming can be expressed in terms of the 2D-DFT of the channel, i.e., $\mathbf{X}$. The SNR at the RX when the TX applies the $(i,j)^{\mathrm{th}}$ element of the 2D-DFT codebook is $\mathrm{SNR}_{\mathrm{BF}}=|X(i,j)|^2/\sigma^2$. The achievable rate corresponding to this SNR is $\mathrm{log}_2(1+\mathrm{SNR}_{\mathrm{BF}})$. For every channel, the exhaustive search-based approach finds the optimal beam index $(i_{\mathrm{opt}},j_{\mathrm{opt}})$ where $|\mathbf{X}|$ achieves its maximum. For the proposed deep learning-based approach, the index corresponding to the predicted beam is used to compute the rate. We compare our algorithm with CS-based beam alignment that uses a random phase shift-based design.
\par The achievable rate plot in Fig. \ref{fig:Rate_vs_SNR} indicates that our deep learning-based approach results in reasonable beam alignment with just $M=10$ channel measurements. Standard CS with a random phase shift-based design, however, requires about $M=40$ channel measurements to achieve comparable performance. It can be observed from Fig. \ref{fig:beam_postquant} that the beams corresponding to our structured CS training focus power in a small set of directions that are more likely to be optimal. Such beams result in a higher SNR in the compressed channel measurements which translates to better beam alignment performance.
\section{Conclusions and future work}
In this paper, we developed a novel approach for compressive beam alignment with millimeter wave phased arrays. Our method is based on a structured compressed sensing technique called 2D-convolutional compressed sensing. Any CS matrix in 2D-CCS can be parameterized by a base matrix and a subsampling set. We showed how deep learning can be used as a tool to optimize the base matrix in 2D-CCS. The optimized base matrix resulted in a phased array compatible structured random CS matrix which achieved superior beam alignment than the common random phase shift-based design. In our future work, we will develop low complexity techniques for beam prediction and also extend our approach to wideband channel estimation with mmWave phased arrays.
\bibliographystyle{IEEEtran}
\bibliography{refs2}
\end{document}